\documentclass[fleqn,10pt,a4paper]{article}
\usepackage{amsmath,amssymb,authblk,bm}
\usepackage{cite,epsfig,color,graphicx,subfigure}
\newcounter{myromancnt}

\newcommand{\xd}{\dot{x}}
\newcommand{\xdd}{\ddot{x}}
\newcommand{\xddd}{\dddot{x}}

\newcommand{\Pref}[1]{(\ref{#1})}

\newcommand{\action}{{\cal S}}

\newcommand{\energy}{{\cal E}}
\newcommand{\V}{{\Phi}}
\newcommand{\mass}{{\cal M}}

\def\dd{{\rm d}}

\title{Self-force on a charged particle in an external scalar field}

\author[1]{Adam Noble}
\author[2]{David A. Burton}
\author[1]{Lauren Docherty}
\author[1]{Dino A. Jaroszynski}
\affil[1]{\small Department of Physics\\ SUPA and University of Strathclyde\\ Glasgow G4 0NG\\ United Kingdom}
\affil[2]{Physics Department\\ Lancaster University\\ Lancaster LA1 4YB\\ United Kingdom}
\begin{document}
\maketitle
\begin{abstract}
A charged particle subject to strong external forces will accelerate, and so radiate energy, inducing a self-force. This phenomenon remains contentious, but advances in laser technology mean we will soon encounter regimes where a more complete understanding is essential. The terms ``self-force'' and ``radiation reaction'' are often used synonymously, but reflect different aspects of the recoil force. For a particle accelerating in an electromagnetic field, radiation reaction is usually the dominant self-force, but in a scalar field this is not the case, and the total effect of the self-force can be anti-frictional. Aspects of this self-force can be recast in terms of spacetime geometry, and this interpretation illuminates the long-standing enigma of a particle radiating while experiencing no self-force.
\end{abstract}

\section{Introduction}
It is well known from classical electrodynamics that an accelerating point charge radiates, and this radiation carries energy and momentum \cite{Jackson1998}. It follows that an additional recoil force (`radiation reaction') must act on the charge, to ensure conservation of the total energy and momentum. In addition, the reconfiguration of the particle's bound field exerts a further force on the charge (`Schott force'). Together, these constitute the particle's self-force, which has led to considerable confusion over the decades. Until recently, this has been a matter of purely theoretical concern, as under most experimental conditions the self-force represented a negligible correction to any applied forces. Rapid advances in laser technology have changed this situation~\cite{Blackburn2020}, and the prospect of experimentally accessing regimes where the self-force is not only non-negligible, but actually dominates the dynamics, has rejuvinated interest in understanding its fundamental properties \cite{Hadad2010,Bulanov2011,Harvey2011,Kravets2013,Capdessus2016,Gonoskov2018,Ekman2021,Ekman2021b}.

The foundations of the electromagnetic self-force were laid by Lorentz \cite{Lorentz1916} and Abraham \cite{Abraham1932}, and made fully relativistic by Dirac \cite{Dirac1938}, using considerations of energy-momentum conservation. The resulting radiation reaction behaves exactly as one would expect, as a frictional force with magnitude corresponding to the Larmor formula for power of emitted radiation. By contrast the Schott force, being proportional to the derivative of acceleration, has no such clear physical interpretation, and notoriously gives rise to unphysical behaviour \cite{Burton2014}.

The equation of motion exhibiting the above behaviour is called the Lorentz-Abraham-Dirac (LAD) equation. That it has no physically acceptable solutions has led many to seek alternative descriptions of the self-force. It has been noted that the need to regularise the infinite bare fields of a point charge leads to an ambiguity in the coefficient of the Schott force \cite{Bhabha1939,Ferris2011}, and this has led some to speculate that it should not feature in the self-force \cite{Bonnor1974,Formanek2020}. For consistency, this requires the particle to lose mass, which is in conflict with observations. Other approaches abandon the point particle assumption \cite{Eliezer1948, Levine1977,Ford1991}, modify Maxwell's equations \cite{Zayats2014,Gratus2015,Kiessling2019}, or introduce {\it ad hoc} descriptions of the self-force \cite{Mo1971,Sokolov2009}. None of these has been universally accepted.

The most widely used description of the self-force is perhaps the most conservative. Introduced by Landau and Lifshitz \cite{Landau1975}, it treats the self-force as a perturbation to the applied force that can be neglected in determining the radiation reaction and Schott forces. This reduces LAD to a second order equation, known as the Landau-Lifshitz (LL) equation, that vanishes in the absence of applied forces, and hence eliminates (most of) the unphysical effects associated with the former, and generally exhibits behaviour in keeping with intuitive expectations, at least in the ultra-relativistic limit.

Although usually applied to situations in which the external forces are electromagnetic, the LAD equation can be used with any external force. By contrast, the LL equation in its standard form requires the applied force to be the Lorentz force due to an external electromagnetic field. There is good reason for this, as in most practical cases of interest the dominant applied force on a charged particle will be electromagnetic, for example from an intense laser pulse or the magnetic fields around astrophysical bodies. But non-electromagnetic forces can also act on a particle. The simplest force one can consider is a scalar interaction. As well as existing at a fundamental level (interactions with the Higgs field, for example, as well as the many scalar fields that emerge in extensions of the Standard Model), these are often used to model more complicated interactions, e.g. the ponderomotive approximation to the Lorentz force of a rapidly oscillating field, or the Yukawa force binding nucleons together in the atomic nucleus. In this paper we explore the form and consequences of the Landau-Lifshitz equation in an applied scalar field (sLL equation), to determine the extent to which the positive features of LL are generic,  rather than being limited to the case of a charged particle accelerating under the Lorentz force.

The remainder of the paper is organised as follows. In Section~\ref{sec:RR} we introduce the mathematical formulation of the LAD and (standard) LL equations, emphasising the roles of the Schott and radiation reaction contributions. In Section~\ref{sec:scalar} we describe a scalar force acting on a charged particle, and determine the form of the sLL equation, before in Section~\ref{sec:conformal} recasting it in terms of a particle radiating in a curved spacetime. In Section~\ref{sec:Noether} we explore how constants of the motion are modified by the self-force, and in Section~\ref{sec:forcefree} we analyse the controversial situation of a particle undergoing uniform proper acceleration, in which the self-force vanishes. Finally, we draw conclusions in Section~\ref{sec:conc}. Throughout the paper we work in Lorentz-Heaviside units, with the speed of light $c=1$. The Einstein summation convention is employed throughout, and indices are raised and lowered with the metric tensor $[\eta_{\mu\nu}]=\text{diag}(1,-1,-1,-1)$.
\section{Electromagnetic self-force}
\label{sec:RR}
Before addressing the self-force on a particle in a scalar field, we recall the basic features of the self-force in general, and in particular its effects on a particle being accelerated by electromagnetic fields.

The story of the self-force on a point charge begins with the LAD equation \cite{Dirac1938},
\begin{equation}
\label{LAD}
\dot{p}^\mu = \underbrace{F^\mu}_\text{Applied force} +\underbrace{\frac{q^2}{6\pi}\xddd^\mu}_\text{Schott force}+\underbrace{\frac{q^2}{6\pi}\xdd^2 \xd^\mu}_\text{Radiation reaction},
\end{equation}
where $x^\mu$ and $p^\mu$ are the worldline and momentum of the particle with charge $q$, $F^\mu$ is an (arbitrary) applied 4-force, and an overdot represents differentiation with respect to proper time $s$.

The self-force terms in \Pref{LAD} comprise two parts, with contrasting behaviours. The `radiation reaction' (RR) contribution\footnote{The term `radiation reaction' is often used to refer to the whole self-force. For our purposes the distinction is important, so we do not here adopt this nomenclature.} behaves exactly as we would expect for radiation friction: it is proportional to the rate of energy emission as given by the Larmor formula, and is oriented antiparallel to the particle's 4-velocity (note $\xdd^2\leq 0$ as acceleration is spacelike). By contrast, the Schott term  (which is all that survives in the nonrelativistic limit) is far less intuitive, and leads to some highly unphysical behaviour. The third derivative generically gives rise to exponentially growing proper acceleration (`runaway solutions'), which can be eliminated only at the cost of having the particle respond to forces that have not yet been applied (`preacceleration'). A further peculiarity of~\Pref{LAD} is that for uniform proper acceleration, $\xdd^2=\text{const.}$, the Schott term precisely cancels the radiation reaction, allowing a situation where a particle radiates without experiencing any consequent recoil force.

Such behaviour is clearly not what one should expect, and a number of suggestions have been made how to resolve these issues. Of the many proposed alternatives \cite{Eliezer1948,Mo1971,Ford1991,Sokolov2009,Hammond2010}, none has met with universal support. Recent experiments have found broad agreement with the predictions of the self-force \cite{Cole2018, Wistisen2018,Poder2018}, albeit in a regime that cannot readily distinguish the different models.

To circumvent these issues, Landau and Lifshitz \cite{Landau1975} introduced an order reduction procedure on \Pref{LAD}, replacing $\xdd^\mu$ and $\xddd^\mu$, respectively, with $m^{-1}F^\mu$ and $m^{-1}\dot{F}^\mu$, where $m$ is the mass of the particle, taking the applied force to be the Lorentz force,
\begin{equation}
\label{Lorentz}
m^{-1}F^\mu=F^\mu{}_\nu\xd^\nu,
\end{equation}
where $F^\mu{}_\nu$ is the electromagnetic field scaled by the particle's charge to mass ratio $q/m$. With the standard relation $p^\mu=m\xd^\mu$, this yields the LL equation,
\begin{equation}
\label{LL}
\xdd^\mu= \underbrace{F^\mu{}_\nu \xd^\nu}_\text{Lorentz force} +\underbrace{\tau \xd^\lambda\partial_\lambda F^\mu{}_\nu \xd^\nu +\tau F^\mu{}_\nu F^\nu{}_\rho \xd^\rho}_\text{Schott force}-\underbrace{\tau F^\lambda{}_\nu F^\nu{}_\rho \xd^\rho\xd_\lambda\xd^\mu}_\text{Radiation reaction},
\end{equation}
where $\tau=q^2/(6\pi m)$ is the `characteristic time' of the particle\footnote{$\tau= 6.2 \times 10^{-24}$ s for an electron or positron, the largest value for any elementary particle.} and terms of order $O(\tau^2)$ are neglected. In contrast to \Pref{LAD}, the LL equation \Pref{LL} is second order, and yields vanishing self-force in regions where $F^\mu{}_\nu=0$, immediately eliminating the runaway and preacceleration problems. Further questions about the reasonableness of LL can be addressed by considering its various contributions to the self-force.

The relative significance of the Schott and RR terms can be seen by denoting the typical field strength (again, scaled by $q/m$) as $\Omega$, its typical frequency as $\omega$, and the particle's relativistic Lorentz factor as $\gamma$. Then the contributions to \Pref{LL} scale relative to the applied force as
\begin{align}
\nonumber \text{Schott (derivative):} \qquad &\frac{\tau \xd\cdot\partial F\cdot\xd}{F\cdot\xd} \sim \tau\omega\gamma, \\
\nonumber \text{Schott (quadratic):} \qquad & \frac{\tau F\cdot F\cdot\xd}{F\cdot\xd} \sim \tau \Omega,\\
\nonumber \text{Radiation reaction:} \qquad &\frac{\tau  (\xd\cdot F\cdot F\cdot\xd)\xd }{F\cdot\xd} \sim \tau \Omega \gamma^2.
\end{align}
Thus the smallness of $\tau$ can be offset by going to high frequencies and energies (in the derivative term) or high field strengths and energies (in the radiation reaction term). It is important to recognise that these are `typical' scalings: particular field configurations can eliminate the enhancement gained by higher energies.

For high energy particles in strong fields (the conditions under which the self-force is important), the radiation reaction term will dominate over the more elusive Schott force, and we can treat the self-force as pure RR. For sufficiently high field strengths and particle energies, it is even possible for the RR force to exceed the applied force. This may appear to violate the assumptions that went into the derivation of \Pref{LL}, but so long as these assumptions are met in the particle's rest frame, the equation remains valid \cite{Bulanov2011,Kravets2013}.

\section{Self-force on a particle interacting with a scalar field}
\label{sec:scalar}

The Landau-Lifshitz equation in the form \Pref{LL} relies on the applied force being the Lorentz force \Pref{Lorentz}. While this is the most obvious force acting on charged particles, it is important to recognise it is not the only possibility. Among the possible types of force that could act on a particle, a scalar force is significant both for its simplicity, and because it can approximate many other interactions (including some cases of the Lorentz force).

Neglecting self-forces, a particle of `scalar charge' $g$, interacting with a scalar field $\V(x)$, is described by the action
\begin{equation}
\label{lagrangian}
\action[x(s)]=-\int \dd s \left( m+g\V\right)\sqrt{\eta_{\mu\nu}\xd^\mu\xd^\nu}.
\end{equation}
If $s$ is proper time, the square root in \Pref{lagrangian} is unity, but we retain it as this parametrisation will not be preserved under variations of the worldline. This action is equivalent to that of a free particle, with position-dependent mass
\begin{equation}
\label{mass}
\mass=m+g\V.
\end{equation}
This is related to the mechanism by which the Higgs field generates fermion masses \cite{Weinberg1967}.

Minimising the action \Pref{lagrangian} leads to the equation of motion
\begin{equation}
\label{momeq}
\dot{p}_\mu= \partial_\mu \mass,
\end{equation}
where the momentum $p_\mu$ is now given by
\begin{equation}
\label{mommass}
p_\mu = \mass \xd_\mu,
\end{equation}
and we have adopted proper time parametrisation. Expressed in terms of acceleration, this corresponds to
\begin{equation}
\label{saccel}
\xdd^\mu=\Delta^\mu{}_\nu f^\nu, \qquad f_\mu=\mass^{-1}\partial_\mu \mass,
\end{equation}
where the $\xd-$orthogonal projection $\Delta^\mu{}_\nu=\delta^\mu{}_\nu-\xd^\mu\xd_\nu$ ensures the proper time condition $\xd^2=1$ is preserved. The field strength $f_\mu$ is analogous to $F_{\mu\nu}$ and the potential $\psi=\ln(\mass/m)$ to $A_\mu$ in the electromagnetic case.

Using \Pref{momeq} as the applied force, and incorporating the LAD self-force, leads to the equation of motion
\begin{equation}
\label{sLAD}
\dot{p}_\mu= \partial_\mu \mass+ \frac{q^2}{6\pi}\left( \xddd_\mu+\xdd^2\xd_\mu\right).
\end{equation}
The appearance of the particle's charge $q$ in (\ref{sLAD}) reminds us that, although the external force is scalar, the last two terms are the self-force due to the particle's {\it electromagnetic field}. While there could be an additional {\it scalar self-force}, its inclusion would merely modify the coefficient in these terms \cite{Barut1975,Johnson2002}.

Following the Landau-Lifshitz order reduction procedure, we replace $\xdd$ and $\xddd$ in~\Pref{sLAD} with \Pref{saccel} and its derivative, yielding
\begin{equation}
\label{sLLRR}
\xdd^\mu=\underbrace{\Delta^\mu{}_\nu f^\nu}_\text{Scalar force}+ \underbrace{\tau_\mass \left[\Delta^\mu{}_\nu \left(\xd^\lambda\partial_\lambda f^\nu-\xd^\lambda f_\lambda f^\nu\right) - f_\lambda \Delta^{\lambda{}\nu} f_\nu\xd^\mu\right]}_\text{Schott force}+ \underbrace{\tau_\mass f_\lambda \Delta^{\lambda{}\nu} f_\nu \xd^\mu}_\text{Radiation reaction},
\end{equation}
where $\tau_\mass=q^2/(6\pi\mass)$ and terms of order $O(\tau^2_\mass)$ are neglected.

The first remark to make about (\ref{sLLRR}) is that the acceleration caused by the self-force is now scaled by the function $\tau_\mass$ rather than the constant $\tau$. This could either enhance or suppress self-force effects, depending on the coupling between the particle and the scalar field. The second, and most crucial, observation regarding this equation is that the last term in the Schott force exactly cancels the radiation reaction, so that the equation of motion reduces to
\begin{equation}
\label{sLL}
\xdd^\mu= \Delta^\mu{}_\nu \left[ f^\nu +\tau_\mass \left( \xd^\lambda\partial_\lambda f^\nu-\xd^\lambda f_\lambda f^\nu\right)\right],
\end{equation}
with no contribution corresponding directly to RR. This does not mean that the particle does not radiate (which any accelerating point charge must, according to the Larmor formula), but rather that the radiation is compensated for by changes to the particle's bound field, rather than by any explicit damping force.

The structure of this scalar Landau-Lifshitz (sLL) equation mirrors that of its electromagnetic counterpart (\ref{LL}), with high energies, strong fields, and rapidly varying fields all enhancing the self-force relative to the applied force. As with the electromagnetic case, it is useful to explore the different scalings of the contributions within the square brackets in \Pref{sLL}:
\begin{align}
\nonumber \text{Derivative term:} \qquad &\frac{\tau_\mass (\xd\cdot\partial) f}{ f} \sim \tau_\mass\omega\gamma, \\
\nonumber \text{Quadratic term:} \qquad &\frac{\tau_\mass (\xd\cdot f) f }{ f} \sim \tau_\mass \Omega \gamma,
\end{align}
where $\omega$ and $\Omega$ are again typical frequencies and magnitudes, respectively, of the field strength $f_\mu$. In this case the stronger energy dependence of \Pref{saccel} compared to the Lorentz force means that the self-force scales with only a single factor of $\gamma$ relative to the applied force, not $\gamma^2$. This also reduces the dominance of the quadratic term ($\xd^\lambda f_\lambda f^\mu$) over the derivative term ($\xd^\lambda\partial_\lambda f^\mu$).

In the ultra-relativistic limit, retaining only the dominant $O(\gamma^3)$ terms, and assuming the field varies slowly so the derivative term can be neglected, (\ref{sLL}) becomes
\begin{equation}
\xdd^\mu \sim \tau_\mass \left( \xd^\nu f_\nu\right)^2\xd^\mu.
\end{equation}
This has exactly the same form as the ultra-relativistic limit of the radiation reaction term in (\ref{sLLRR}), but the opposite sign. In this limit, the total self-force acts as {\it radiation anti-damping}.

As a final remark, we note that the self-force can be elegantly recast in terms of the effective mass $\mass$ and momentum $p_\mu$ in \Pref{mass} and \Pref{mommass}:
\begin{equation}
\label{sLLp}
 \dot{p}_\mu= \partial_\mu\mass -\frac{q^2}{6\pi} \mass\Delta^\nu{}_\mu \xd^\lambda\partial_\lambda\partial_\nu \mass^{-1}.
\end{equation}
This form of \Pref{sLL} will prove convenient in analysing the physical consequences of the self-force in a scalar field, to which we will turn in Section~\ref{sec:Noether}. Before doing so, we explore how a scalar force can be reinterpreted geometrically, and how the self-force appears in this interpretation.

\section{Radiating particle in a conformally flat spacetime}
\label{sec:conformal}

The interaction of a particle with a scalar field admits an alternative interpretation in terms of a curved spacetime. Absorbing the dynamical mass $\mass=m+g\Phi$ into the metric inside the square root, the action (\ref{lagrangian}) is seen to be equivalent to that for a free particle in a conformally flat spacetime,
\begin{equation}
\label{conformal}
\action[x(s)]=-m\int \dd s \sqrt{g_{\mu\nu}\xd^\mu\xd^\nu}, \qquad g_{\mu\nu}=e^{2\psi} \eta_{\mu\nu},
\end{equation}
where the conformal factor $\psi=\ln(\mass/m)$ is the scalar potential, as above\footnote{The scalar potential depends on the particle's scalar charge to mass ratio $g/m$, so different particle species will experience different effective metrics.}. Variation of the action (\ref{conformal}) yields the geodesic equation
\begin{equation}
\label{geodesic}
a^\mu= \frac{\dd u^\mu}{\dd\lambda}+\Gamma^\mu{}_{\nu\rho}u^\nu u^\rho=0,
\end{equation}
where the connection coefficients are given by
\begin{equation}
\Gamma^\mu{}_{\nu\rho}=f_\nu\delta^\mu_\rho+f_\rho\delta^\mu_\nu-\eta^{\mu\alpha}f_\alpha \eta_{\nu\rho},
\end{equation}
with $f_\mu=\partial_\mu\psi$. $\lambda$ is proper time with respect to the metric $g_{\mu\nu}$, and the 4-velocity $u^\mu$ satisfies the normalisation $g_{\mu\nu}u^\mu u^\nu=1$.

The equation of motion (\ref{saccel}) can be recovered from (\ref{geodesic}) by substituting $\xd^\mu=e^\psi u^\mu$ and changing parametrisation to $s$, proper time with respect to the flat metric $\eta_{\mu\nu}$, using $\dd s=e^{-\psi} \dd\lambda$. This demonstrates the equivalence of the action of a scalar force in flat spacetime and geodesic motion in a curved spacetime for a {\it free particle}.

This interpretation can be extended to include the self-forces, thanks to the conformal invariance of Maxwell's equations. In a curved spacetime, the LAD equation~(\ref{LAD}) generalises to\footnote{We could add an external force to the RHS of (\ref{conformalLAD}). Since the scalar force of interest has been absorbed into the spacetime geometry, we omit this term.} \cite{Dewitt1960,Hobbs1968a}
\begin{equation}
\label{conformalLAD}
a^\mu= \tau\Delta^\mu{}_\nu \left(\frac{{\rm D}a^\nu}{\dd\lambda} - \frac{1}{2} R^\nu{}_\lambda u^\lambda\right)+ 3\tau u_\nu \int \left( \nabla^\mu G^\nu{}_\lambda- \nabla^\nu G^\mu{}_\lambda \right) u^\lambda \dd s^\prime,
\end{equation}
where $\nabla_\mu$ is the covariant derivative and ${\rm D}/\dd\lambda=u^\mu \nabla_\mu$ the absolute derivative, and $G^\mu{}_\nu$ is the retarded Green's function for Maxwell's equations, all in the spacetime with metric $g_{\mu\nu}$. For convenience, in (\ref{conformalLAD}) we raise and lower indices with $g_{\mu\nu}$ and its inverse (in all other equations in this article, index raising and lowering is performed with the flat metric $\eta_{\mu\nu}$).

The final term in (\ref{conformalLAD}) is nonlocal, depending on the past history of the particle's worldline. However, in conformally flat spacetime this tail vanishes \cite{Hobbs1968b,Roberts1989}, leaving a local equation. With the conformally flat metric, the Ricci tensor takes the form
\begin{equation}
\label{Ricci}
R_{\mu\nu}=-2\left(\partial_\mu f_\nu- f_\mu f_\nu\right) -\eta^{\alpha\beta}\left( \partial_\alpha f_\beta+2 f_\alpha f_\beta\right)\eta_{\mu\nu}.
\end{equation}

Substituting (\ref{Ricci}) into (\ref{conformalLAD}), and again converting from $g_{\mu\nu}$, $u^\mu$ and $\lambda$ to $\eta_{\mu\nu}$, $\xd^\mu$ and $s$, yields
\begin{equation}
\label{conformalLL}
\xdd^\mu= \Delta^{\mu\nu}\left[ f_\nu+ \tau_\mass \left( \xd^\lambda\partial_\lambda f_\nu - \xd^\lambda f_\lambda f_\nu +\frac{{\rm D}a_\nu}{\dd s}\right)\right].
\end{equation}

This equation is almost identical to (\ref{sLL}), differing only in the final term involving the derivative of the acceleration. This distinction serves to remind us that (\ref{conformalLL}) is the curved spacetime equivalent to the LAD equation (\ref{LAD}), and has not yet been subjected to the Landau-Lifshitz order reduction procedure. As such, the similarity between the two equations is quite remarkable. The equivalence can be made complete by recognising that the acceleration $a^\mu$ vanishes in the absence of self-forces (recall from \Pref{geodesic} that the effects of the scalar field are here encoded in the connection coefficients $\Gamma^\mu{}_{\nu\rho}$), so the order reduction merely eliminates the final term in (\ref{conformalLL}).

We conclude this Section by remarking that this identification of self-forces on a particle accelerated by a scalar force,  with those on an otherwise free particle in curved spacetimes, goes both ways. Thus the results of the current work may apply more generally than to particles under the influence of scalar forces, finding use for example in the early universe, with the scalar force mimicking the effects of the cosmological scale factor.

\section{Evolution of Noether charges}
\label{sec:Noether}

Noether's theorem tells us that to every symmetry of the action (in the present context, every symmetry of $\mass$) there corresponds a conserved quantity \cite{Noether1918}. However, inclusion of the self-force violates the conditions of Noether's theorem, giving rise to the possibility that these `conserved quantities' can change in time. Noether charges are important for determining the integrability of dynamics in a scalar field \cite{Ansell2018}, and understanding the rate at which they evolve is an important starting point to verifying theories of the self-force. 

In this Section we consider fields exhibiting symmetries, and determine the evolution of their corresponding Noether charges. We focus on the two examples of $\mass$ that are invariant under time translation, and under spatial translations, both because these are in a sense the most fundamental cases, and also because they give the clearest insight into the effects of self-force. The extension to other Poincar\'{e} symmetries is straightforward.

\subsection{$\mass({\vec{r}})$ and the variation of energy}

An action that is invariant in time yields the total energy $\energy$ as a conserved quantity. From \Pref{lagrangian}, this energy has the form
\begin{equation}
\energy= p_0 = \mass \gamma.
\end{equation}
This could be separated into kinetic ($m\gamma$) and potential ($g\V\gamma$) energies, though since $\mass$ is the measurable mass, such a distinction is not particularly useful.

As expected, \Pref{saccel} yields $\energy=\text{const}$. By contrast, the self-force equation \Pref{sLL} or \Pref{sLLp} yields a change of energy at the rate
\begin{equation}
\frac{\dd\cal E}{\dd t}=\frac{q^2}{6\pi}\mass (\vec{u}\cdot\vec{\nabla})^2 \mass^{-1}.
\end{equation}
Note that we treat position $\vec{r}$ and velocity $\vec{u}=\dd\vec{r}/\dd s$ as independent variables, so $\vec{\nabla}$ acts only on $\mass^{-1}$, and not on $\vec{u}$.

It is clear that $\energy$ may increase or decrease, depending on how $\mass$ varies in the direction of the particle's motion. This emphasises that the self-force cannot be interpreted as purely a radiation reaction force, as the radiation can only reduce the total particle energy. This is in contrast to the analogous case of LL in an applied electrostatic field, where an ultra-relativistic particle will typically lose energy\footnote{In the Sokolov model of the self-force \cite{Sokolov2009}, the particle can {\it only} lose energy in an electrostatic field, regardless of whether or not it is ultra-relativistic \cite{Capdessus2016}.}.

\subsection{$\mass(t)$ and the variation of momentum}

If $\mass$ is a function of time only, the (spatial) momentum $\vec{p}=\mass\vec{u}$ would be conserved, according to Noether's theorem. When the self-force is taken into account this is no longer the case, and instead \Pref{sLL} yields
\begin{equation}
\label{pevolution}
\frac{\dd\vec{p}}{\dd t}=\frac{q^2}{6\pi}  \frac{\dd^2\mass^{-1}}{\dd t^2}\gamma \vec{p}.
\end{equation}
In this case, the self-force acts to change the {\it magnitude} of momentum, but not its direction. This is indeed what we would expect of a frictional force, except that depending on how $\mass$ is changing in time, the momentum could increase rather than decrease.

As an illustration, consider a spatially flat radiation-dominated Friedmann-Lema\^itre-Robertson-Walker spacetime, modelling the early universe. In terms of {\it conformal time} the metric of such a spacetime is written
\begin{equation}
g_{\mu\nu}= \alpha^2 t^2 \eta_{\mu\nu}.
\end{equation}
The equivalence between scalar fields and conformally flat spacetime allows us to interpret this as a scalar field, giving particles an effective mass
\begin{equation}
\mass= m\alpha t,
\end{equation}
and hence (\ref{pevolution}) yields
\begin{equation}
\frac{\dd\vec{p}}{\dd t}=2\tau_\mass \frac{\gamma}{t^2}\vec{p},
\end{equation}
implying that in the very early universe, charged particles may obtain a greater momentum than would be suggested from considerations neglecting the self-force.

\section{Self-force--free motion}
\label{sec:forcefree}

We end this exploration of the self-force in a scalar field by revisting a notorious apparent paradox of the LAD equation, which unlike the problems of runaway solutions and preacceleration is inherited by LL. A particle undergoing hyperbolic motion satisfies the relation $\xddd^\mu={\rm a}^2\xd^\mu$, where ${\rm a}=\sqrt{-\xdd^2}$ is the (constant) proper acceleration. Hence the Schott term in \Pref{LAD} exactly cancels the radiation reaction force, and the total self-force vanishes.

This combination of nonzero acceleration and vanishing self-force has caused considerable confusion over the years, and led to a long debate over whether or not a charge in hyperbolic motion would radiate \cite{Fulton1960}. This apparent paradox has now largely been resolved, with the explanation that the radiated energy is `borrowed' from the particle's Coulomb field, and the balance is paid off with rapid changes in the latter in the transitions into and out of the period of uniform acceleration \cite{Steane2015b}.

While this explanation is generally accepted, it is not entirely satisfactory that it relies on the behaviour outside the period of hyperbolic motion to fully account for energy-momentum conservation. As such, uncovering new aspects of self-forces in hyperbolic motion remains an active field of research \cite{Eriksen2000,Eriksen2002,Eriksen2004,Steane2015,Kang2021}.

From \Pref{sLLp} it is clear that the self-force vanishes for fields of the form
\begin{equation}
\label{sffree}
\mass=\frac{m}{1-b_\mu x^\mu},
\end{equation}
where $b_\mu$ is any constant vector, and such fields do indeed accelerate particles at a constant proper rate. The first point to note is that this field diverges as the particle approaches the point $b_\mu x^\mu=1$, so that the period of uniform acceleration cannot be continued indefinitely.

We can gain further insight by switching to the curved spacetime interpretation of the scalar field. From (\ref{Ricci}) and the expression for $f_\mu$ in (\ref{saccel}), the field (\ref{sffree}) corresponds to a Ricci tensor of the form
\begin{equation}
R_{\mu\nu}= -3 \frac{b^\alpha b_\alpha}{(1-b_\beta x^\beta)^2}\eta_{\mu\nu}.
\end{equation}
Of particular interest is the case where $b_\mu$ is a null vector, $b^\mu b_\mu=0$. In this case the Ricci tensor vanishes which, together with conformal flatness, implies that  $g_{\mu\nu}$ is {\it identically flat} (albeit expressed in non-inertial coordinates). Thus the equivalence between the motion of a point particle due to a scalar field and the geometrical description of the motion of a point particle in a conformally related spacetime shows that hyperbolic motion can be interpreted as geodesic motion in an alternative {\it flat spacetime}, for which self-forces would naturally be expected to vanish.

\section{Conclusions}
\label{sec:conc}

The behaviour of a particle under the influence of some external force is one of the most basic questions in physics, and yet one that is greatly obscured by the presence of the self-force. A charged particle is immersed not only in any external fields present, but also in the electromagnetic field it generates. The response to the particle's own field is the cause of considerable confusion, and for over a century has caused problems of interpretation. 

The most widely recognised contribution to a particle's self-force is radiation reaction, to the extent that this label is often applied to the entire self-force. In the ultra-relativistic limit, radiation is emitted in a highly collimated beam in the direction of the particle's motion \cite{Jackson1998}. As such, radiation reaction has a frictional nature. Deviations from this behaviour arise from the other contribution to the self-force, the Schott force, generated by changes to the particle's bound fields.

When the external field is electromagnetic, the self-force is the well-known Landau-Lifshitz force \cite{Landau1975}. As expected, in the ultra-relativistic limit this self-force is typically frictional, with magnitude given by the Larmor formula. The situation is very different when the external field is a scalar. Here, the radiation reaction is precisely cancelled by part of the Schott force, so that the remaining self-force is not frictional, even in the ultra-relativistic limit, and can even be anti-damping. This leads to behaviour that would not be expected from radiation reaction alone, such as a particle in a time-independent field gaining energy.

The action of a scalar force on a test particle can be interpreted in terms of a free particle in a conformally flat spacetime, and this equivalence can be extended to include the self-force. This expands the applicability of the current work, for example to include certain cosmological models. It also allows for a new interpretation of the old problem of hyperbolic motion without self-force, which can be regarded as equivalent to inertial motion in a flat spacetime.

\section*{Acknowledgements}
This work was supported by the UK EPSRC (Grant No. EP/N028694/1) and the European Union H2020 Research and Innovation Programme LASERLAB EUROPE (Grant No. 871124). AN is also grateful for support from the Cockcroft Institute. All of the results can be fully reproduced using the methods described in the paper.


\begin{thebibliography}{99}
\bibitem{Jackson1998}
Jackson J~D 1998 {\it Classical Electrodynamics\/} 3rd ed (New York: Wiley)

\bibitem{Blackburn2020}
Blackburn T 2020 Radiation reaction in electron-beam interactions with high-intensity lasers {\it Rev. Mod. Plasma Phys.\/} {\bf 4} 5

\bibitem{Hadad2010}
Hadad Y, Labun L, Rafelski J, Elkina N, Klier C and Ruhl H 2010 Effects of radiation reaction in relativistic laser acceleration {\it Phys. Rev.
  D\/} {\bf 82} 096012

\bibitem{Bulanov2011}
Bulanov S~V, Esirkepov T~Z, Kando M, Koga J~K and Bulanov S~S 2011 Lorentz-Abraham-Dirac versus Landau-Lifshitz radiation friction force in the ultrarelativistic electron interaction with electromagnetic wave (exact solutions) {\it Phys.
  Rev. E\/} {\bf 84} 056605

\bibitem{Harvey2011}
Harvey C, Heinzl T and Marklund M 2011 Symmetry breaking from radiation reaction in ultra-intense laser fields {\it Phys. Rev. D\/} {\bf 84} 116005

\bibitem{Kravets2013}
Kravets Y, Noble A and Jaroszynski D~A 2013 Radiation reaction effects on the interaction of an electron with an intense laser pulse {\it Phys. Rev. E\/} {\bf 88}
  011201(R)

\bibitem{Capdessus2016}
Capdessus R, Noble A, McKenna P and Jaroszynski D~A 2016 Role of momentum and velocity for radiating electrons {\it Phys. Rev. D\/}
  {\bf 93} 045034

\bibitem{Gonoskov2018}
Gonoskov A and Marklund M 2018 Radiation-dominated particle and plasma dynamics {\it Physics of Plasmas\/} {\bf 25} 093109
 
\bibitem{Ekman2021}
Ekman R, Heinzl T and Ilderton A 2021 Exact solutions in radiation reaction and the radiation-free direction {\it New J. Phys.\/} {\bf 23}
  055001 

\bibitem{Ekman2021b}
Ekman R, Heinzl T and Ilderton A 2021 Reduction of order, resummation, and radiation reaction {\it Phys. Rev. D\/} {\bf 104} 036002
 

\bibitem{Lorentz1916}
Lorentz H~A 1916 {\it The Theory of Electrons and its Applications to the
  Phenomena of Light and Radiant Heat\/} 1st ed (New York: Stechert)

\bibitem{Abraham1932}
Abraham M 1932 {\it The Classical Theory of Electricity and Magnetism\/} 1st ed
  (London: Blackie)

\bibitem{Dirac1938}
Dirac P~A~M 1938 Classical Theory of Radiating Electrons {\it Proc. R. Soc. Lond. A\/} {\bf 167} 148--169

\bibitem{Burton2014}
Burton D~A and Noble A 2014 Aspects of electromagnetic radiation reaction in strong fields {\it Contemporary Physics\/} {\bf 55} 110--121

\bibitem{Bhabha1939}
Bhabha H~J 1939 Classical theory of mesons {\it Proc. R. Soc. Lond. A\/} {\bf 172} 384--409

\bibitem{Ferris2011}
Ferris M~R and Gratus J 2011 The origin of the Schott term in the electromagnetic self force of a classical point charge {\it J. Math. Phys.\/} {\bf 52} 092902

\bibitem{Bonnor1974}
Bonnor W~B 1974 A new equation of motion for a radiating charged particle {\it Proc. R. Soc. Lond. A\/} {\bf 337} 591--598

\bibitem{Formanek2020}
Formanek M, Steinmetz A and Rafelski J 2020 Radiation reaction friction: Resistive material medium {\it Phys. Rev. D\/} {\bf 102}
  056015

\bibitem{Eliezer1948}
Eliezer C~J 1948 On the classical theory of particles {\it Proc. R. Soc. Lond. A\/} {\bf 194} 543--555

\bibitem{Levine1977}
Levine H, Moniz E~J and Sharp D~H 1977 Motion of extended charges in classical electrodynamics {\it Am. J. Phys.\/} {\bf 45} 75--78

\bibitem{Ford1991}
Ford G W and O'Connell R F 1991 Radiation reaction in electrodynamics and the elimination of runaway solutions {\it Physics Letters A\/} {\bf 157} 217--220

\bibitem{Zayats2014}
Zayats A~E 2014 Self-interaction in the Bopp-Podolsky electrodynamics: Can the observable mass of a charged particle depend on its acceleration? {\it Annals of Physics\/} {\bf 342} 11--20 

\bibitem{Gratus2015}
Gratus J, Perlick V and Tucker R~W 2015 On the self-force in Bopp-Podolsky electrodynamics {\it J. Phys. A\/} {\bf 48} 435401

\bibitem{Kiessling2019}
Kiessling M~K-H 2019 Force on a point charge source of the classical electromagnetic field {\it Phys. Rev. D\/} {\bf 100} 065012

\bibitem{Mo1971}
Mo T~C and Papas C~H 1971 New Equation of Motion for Classical Charged Particles {\it Phys. Rev. D\/} {\bf 4} 3566--3571

\bibitem{Sokolov2009}
Sokolov I~V 2009 Renormalization of the Lorentz-Abraham-Dirac equation for radiation reaction force in classical electrodynamics {\it JETP\/} {\bf 16} 207--212

\bibitem{Landau1975}
Landau L~D and Lifshitz E~M 1975 {\it The Classical Theory of Fields\/} 4th ed
  vol~2 (Oxford: Butterworth-Heinemann)

\bibitem{Hammond2010}
Hammond R~T 2010 Radiation reaction at ultrahigh intensities {\it Phys. Rev. A\/} {\bf 81} 062104

\bibitem{Cole2018}
Cole J~M, Behm K~T, Gerstmayr E, Blackburn T~G, Wood J~C, Baird C~D, Duff M~J,
  Harvey C, Ilderton A, Joglekar A~S, Krushelnick K, Kuschel S, Marklund M,
  McKenna P, Murphy C~D, Poder K, Ridgers C~P, Samarin G~M, Sarri G, Symes D~R,
  Thomas A~G~R, Warwick J, Zepf M, Najmudin Z and Mangles S~P~D 2018 Experimental  evidence  of  radiation  reaction  in  the  collision  of  a  high-intensity laser pulse with a laser-wakefield accelerated electron beam {\it Phys.
  Rev. X\/} {\bf 8} 011020

\bibitem{Wistisen2018}
Wistisen T~N, Di~Piazza A, Knudsen H~V and Uggerh{\o}j U~I 2018 Experimental evidence of quantum radiation reaction in aligned crystals {\it Nat.  Commun.\/} {\bf 9} 795

\bibitem{Poder2018}
Poder K, Tamburini M, Sarri G, Di~Piazza A, Kuschel S, Baird C~D, Behm K,
  Bohlen S, Cole J~M, Corvan D~J, Duff M, Gerstmayr E, Keitel C~H, Krushelnick
  K, Mangles S~P~D, McKenna P, Murphy C~D, Najmudin Z, Ridgers C~P, Samarin
  G~M, Symes D~R, Thomas A~G~R, Warwick J and Zepf M 2018 Experimental signatures of the quantum nature of radiation reaction in the field of an ultraintense laser {\it Phys. Rev. X\/}
  {\bf 8} 031004

\bibitem{Weinberg1967}
Weinberg S 1967 A Model of Leptons {\it Phys. Rev. Lett.\/} {\bf 19} 1264

\bibitem{Barut1975}
Barut A~O and Villarroel D 1975 Radiation reaction amd mass renormalization in scalar and tensor fields and linearized gravitation {\it J. Phys. A: Math. Gen.\/} {\bf 8} 156--159

\bibitem{Johnson2002}
Johnson P~R and Hu B~L 2002 Stochastic theory of relativistic particles moving in a quantum field: Scalar Abraham-Lorentz-Dirac-Langevin equation, radiation reaction, and vacuum fluctuations {\it Phys. Rev. D\/} {\bf 65} 065015

\bibitem{Dewitt1960}
DeWitt B~S and Brehme R~W 1960 Radiation damping in a gravitational field {\it Annals of Physics\/} {\bf 9} 220--259 

\bibitem{Hobbs1968a}
Hobbs J 1968 A vierbein formalism of radiation damping {\it Annals of Physics\/} {\bf 47} 141--165 

\bibitem{Hobbs1968b}
Hobbs J 1968 Radiation damping in conformally flat universes {\it Annals of Physics\/} {\bf 47} 166--172 

\bibitem{Roberts1989}
Roberts M~D 1989 The motion of a charged particle in a spacetime with a conformal metric {\it Class. Quantum Grav.\/} {\bf 6} 419--423

\bibitem{Noether1918}
Noether E 1918 Invariante Variationsprobleme {\it Nachr. d. K\"{o}nig. Gesellsch. d. Wiss. zu G\"{o}ttingen,
  Math-phys. Klasse\/} {\bf 1} 235--257

\bibitem{Ansell2018}
Ansell L, Heinzl T and Ilderton A 2018 Superintegrable relativistic systems in scalar background fields {\it J. Phys. A: Math. Theor.\/} {\bf
  51} 495203

\bibitem{Fulton1960}
Fulton T and Rohrlich F 1960 Classical radiation from a uniformly accelerated charge {\it Annals of Physics\/} {\bf 9} 499--517

\bibitem{Steane2015b}
Steane A~M 2015 Tracking the radiation reaction energy when charged bodies accelerate {\it Am. J. Phys.\/} {\bf 83} 703--710

\bibitem{Eriksen2000}
Eriksen E and Gr{\o}n {\O} 2000 Electrodynamics of Hyperbolically Accelerated Charges: I. The Electromagnetic Field of a Charged Particle with Hyperbolic Motion {\it Annals of Physics\/} {\bf 286} 320--342

\bibitem{Eriksen2002}
Eriksen E and Gr{\o}n {\O} 2002 Electrodynamics of Hyperbolically Accelerated Charges: IV. Energy-Momentum Conservation of Radiating Charged Particles {\it Annals of Physics\/} {\bf 297} 243--294
 
\bibitem{Eriksen2004}
Eriksen E and Gr{\o}n {\O} 2004 Electrodynamics of hyperbolically accelerated charges V. The field of a charge in the Rindler space and the Milne space {\it Annals of Physics\/} {\bf 313} 147--196

\bibitem{Steane2015}
Steane A~M 2015 Self-force of a rigid ideal fluid, and a charged sphere in hyperbolic motion {\it Phys. Rev. D\/} {\bf 91} 065008

\bibitem{Kang2021}
Kang T, Noble A, Yoffe S~R, Jaroszynski D~A and Hur M~S 2021 Radiation reaction and the acceleration-dependent mass increase of a charged sphere undergoing uniform acceleration {\it Phys.
  Lett. A\/} {\bf 407} 127445
\end{thebibliography}
\end{document}